\documentclass[
aps,
pra,
preprint,
superscriptaddress,
showkeys,
floatfix
]{revtex4-2}

\usepackage[utf8]{inputenc}
\usepackage[T1]{fontenc}

\usepackage{graphicx,xspace}
\usepackage{amssymb}
\usepackage{mathtools}
\usepackage{booktabs}
\usepackage{multirow}
\usepackage[margin=2.5cm]{geometry}
\usepackage{lineno}

\usepackage{amsmath}
\usepackage{tikz}

\newcommand\datasetname{MP-ALOE}
\newcommand\papername{\datasetname: An r$^2$SCAN dataset for universal machine learning interatomic potentials\xspace}

\newcommand\angstrom{\r{A}}

\usepackage{xcolor, soul}
\sethlcolor{green}

\usepackage{hyperref}

\usepackage{titlesec}
\titlespacing*{\section}{0pt}{*}{*}

\titleformat{\section}
  {\normalfont\Large\bfseries}
  {\thesection}{1em}{}

\titleformat{\subsection}
  {\normalfont\large\bfseries}
  {\thesubsection}{1em}{}

\titleformat{\subsubsection}
  {\normalfont\bfseries}
  {\thesubsubsection}{1em}{}

\usepackage{caption}
\usepackage{setspace}
\DeclareCaptionFont{smalllinespacing}{\small\setstretch{0.9}}

\begin{document}

\title{\papername}

\author{Matthew C. Kuner}
\email[]{matthewkuner@berkeley.edu}
\affiliation{Materials Science Division, Lawrence Berkeley National Laboratory, Berkeley, California 94720, USA}
\affiliation{Department of Materials Science and Engineering, University of California, Berkeley, California 94720, USA}

\author{Aaron D. Kaplan}
\affiliation{Materials Science Division, Lawrence Berkeley National Laboratory, Berkeley, California 94720, USA}

\author{Kristin A. Persson}
\affiliation{Materials Science Division, Lawrence Berkeley National Laboratory, Berkeley, California 94720, USA}
\affiliation{Department of Materials Science and Engineering, University of California, Berkeley, California 94720, USA}

\author{Mark Asta}
\affiliation{Materials Science Division, Lawrence Berkeley National Laboratory, Berkeley, California 94720, USA}
\affiliation{Department of Materials Science and Engineering, University of California, Berkeley, California 94720, USA}

\author{Daryl C. Chrzan }
\email[]{dcchrzan@berkeley.edu}
\affiliation{Materials Science Division, Lawrence Berkeley National Laboratory, Berkeley, California 94720, USA}
\affiliation{Department of Materials Science and Engineering, University of California, Berkeley, California 94720, USA}

\begin{abstract}
We present \datasetname, a dataset of nearly 1 million DFT calculations using the accurate r$^2$SCAN meta-generalized gradient approximation. Covering 89 elements, \datasetname\, was created using active learning and primarily consists of off-equilibrium structures. We benchmark a machine learning interatomic potential trained on \datasetname, and evaluate its performance on a series of benchmarks, including predicting the thermochemical properties of equilibrium structures; predicting forces of far-from-equilibrium structures; maintaining physical soundness under static extreme deformations; and molecular dynamic stability under extreme temperatures and pressures. \datasetname\, shows strong performance on all of these benchmarks, and is made public for the broader community to utilize. 
\end{abstract}



\maketitle

\clearpage
\section*{Introduction}
\label{sec:Introduction}

Atomistic simulations are among the most common tools used by computational materials scientists. While \textit{ab initio} density functional theory (DFT) \cite{kohn_self-consistent_1965} simulations show impressive accuracy to experiment due to adherence to physical constraints \cite{kaplan_predictive_2023, tran_rungs_2016}, they are often quite expensive. For solid systems specifically, modern plane-wave DFT \cite{blochl_projector_1994} typically limits users to simulating, at most, hundreds of atoms and timescales on the order of picoseconds. Hence, more efficient methods are needed for calculating quantities which are numerically intensive (e.g., transition state searches) or which require large simulation cells (e.g., diffusion pathways or equilibration of non-crystalline materials). 

Classical, empirical force-fields for solids, such as the (modified) embedded atom method \cite{daw_embedded-atom_1984, baskes_modified_1992}, allow users to perform dynamic simulations at low cost. However, the fixed functional forms of such classical force-fields limit their accuracy and generalizability to narrow chemical regimes.
Additionally, there are no classical forcefields for solids which span almost the entire periodic table, nor which can adequately treat stretched bonds encountered in solid-state reactions.
In recent years, machine-learning interatomic potentials (MLIPs) \cite{behler_2007_nnip} have increasingly become a promising alternative. Such potentials allow a researcher to curate a set of \textit{ab initio} calculations to train an MLIP, which can approximate the potential energy surface (PES) based on the training set alone. MLIPs are often implemented as Graph Neural Networks (GNNs), given that graphs are a natural way to represent the chemical bonding of atoms. Notable examples include Refs. \cite{gilmer_neural_2017, schutt_schnet_2017, batzner_e3-equivariant_2022, drautz_atomic_2019, batatia_mace_2022}.

The apex of MLIP research would be to create a universal MLIP (UMLIP) that can accurately approximate a given DFT functional across the periodic table. 
The current generation of UMLIPs, beginning with M3GNet \cite{chen_universal_2022} and continuing with, e.g., Refs. \cite{chen_universal_2022,deng_chgnet_2023,batatia_foundation_2023,park_scalable_2024,yang_mattersim_2024,barroso-luque_open_2024} cover most of the periodic table (at least 89 elements) and maintain close-to-linear scaling with the number of simulated atoms, exponentially faster than DFT's cubic scaling with the number of \emph{electrons}. 
However, modern pre-trained models are typically limited in accuracy and transferability: models trained primarily to equilibrium solid-state data tend to underestimate energy and forces in out-of-domain test cases \cite{deng_systematic_2025}. 
Further, the accuracy of smaller models tends to exceed the uncertainty of experiment in predicting equilibrium compositional phase stability \cite{riebesell_matbench_2024}, whereas much larger models are less computationally tractable.

One key challenge in improving UMLIPs is increasing the quality of the underlying DFT data they are trained on. Most current UMLIPs are trained on DFT calculations at the Perdew-Burke-Ernzerhof (PBE) generalized gradient approximation (GGA) level of theory \cite{perdew_generalized_1996}, sourced from the Materials Project (MP) database \cite{jain_commentary_2013, horton_accelerated_2025}, the Alexandria database \cite{schmidt_crystal_2021, schmidt_machine-learning-assisted_2023}, and/or the OMat24 dataset \cite{barroso-luque_open_2024}. 
While PBE is often accurate for describing simple $sp$-bonded solids, it struggles to describe weaker bonds present in mixed-/ionic and dispersion-bound solids \cite{tran_rungs_2016}, as well as systems plagued by delocalization errors \cite{kaplan_predictive_2023}, such as defects \cite{nazarov_vacancy_20212}.

To date, only one solid-state dataset \cite{kaplan_foundational_2025} has been calculated at a higher level of approximation, using the r$^2$SCAN meta-GGA \cite{furness_accurate_2020}.
Meta-GGAs typically improve systematically over GGAs like PBE \cite{kingsbury_performance_2022, kothakonda_testing_2023, swathilakshmi_performance_2023}, and often perform comparably to or better than the higher hybrid levels of theory for calculation of equilibrium solid-state properties \cite{liu_hybrid_2024,kothakonda_testing_2023}.

The MatPES dataset of Ref. \cite{kaplan_foundational_2025} -- the only public r$^2$SCAN dataset for UMLIPs released before this work -- showed strong results on a series of near- and off-equilibrium benchmarks, especially given its smaller size relative to other datasets. However, MatPES is still limited to relatively low-energy structures sampled from 300K molecular dynamics (MD) trajectories. Moreover, the chemical compositions (and initial structures for MD) included are exclusively sourced from the compounds in the Materials Project. Although MatPES reflects the chemical environments of the Materials Project,
MatPES-trained UMLIPs were able to reasonably capture the formation energies of solids whose compositions were not present in the dataset.

MatPES demonstrably samples a wider distribution of interatomic forces and stresses than the Materials Project relaxation trajectories, partly attributable to its source of structures (MD) and partly to its sampling method \cite{qi_robust_2024} which attempts to increase dissimilarity of the sampled structures.
It is currently unclear how wide a force distribution is needed to ensure general stability for MD, and to stiffen the PES at larger interatomic separations \cite{deng_systematic_2025}.
Approaches such as active learning, applied either to training potentials in limited chemical regime \cite{lysogorskiy_active_2023} or to the whole periodic table \cite{yang_mattersim_2024}, can aid in systematically improving coverage of PES regions where a UMLIP has a low density of training data.
Active learning approaches for UMLIPs currently rely on minimizing model uncertainty either via mathematical quantities derived from the model architecture \cite{lysogorskiy_active_2023} or from physical property-motivated uncertainty estimators \cite{yang_mattersim_2024}.

In this work, we present a new r$^2$SCAN dataset for UMLIPs which begins to address some of the previously mentioned open questions. This dataset was created via elemental substitution of prototype structures, and augmented using active learning (AL) via query by committee (QBC) \cite{seung_query_1992}. We have named it \datasetname\, (\textbf{M}aterials \textbf{P}roject - \textbf{A}ctive \textbf{L}earning of \textbf{O}ff \textbf{E}quilibrium structures). The \datasetname\, dataset contains a greater sampling of high energy structures, large magnitude forces, and high pressures than in MatPES. Moreover, given that \datasetname\, and MatPES were calculated using compatible DFT settings, we then perform a series of benchmarks comparing MACE \cite{batatia_mace_2022} models trained on \datasetname, MatPES, and both of these datasets combined. Specifically, we look at predictions for equilibrium properties, off-equilibrium forces, static deformations under extreme hydrostatic pressure, and MD stability under extreme ensemble conditions. We find \datasetname- and MatPES-trained MACE models to be competitive in predicting equilibrium energies and off-equilibrium forces. 
Notably, the \datasetname-trained model demonstrates improved stability in MD runs and physicality of the PES under static extreme hydrostatic pressures.
By construction, \datasetname\, is completely compatible with the r$^2$SCAN subset of MatPES; a MACE model trained on their union clearly demonstrates the strongest overall performance.

\section*{Results}
\label{sec:Results}

\subsection*{The Dataset}
\label{subsec:Dataset}

\begin{figure}[htb]
	\makebox[\textwidth][c]{\includegraphics[width=1.0\linewidth,trim={0mm 0mm 0mm 0mm},clip]{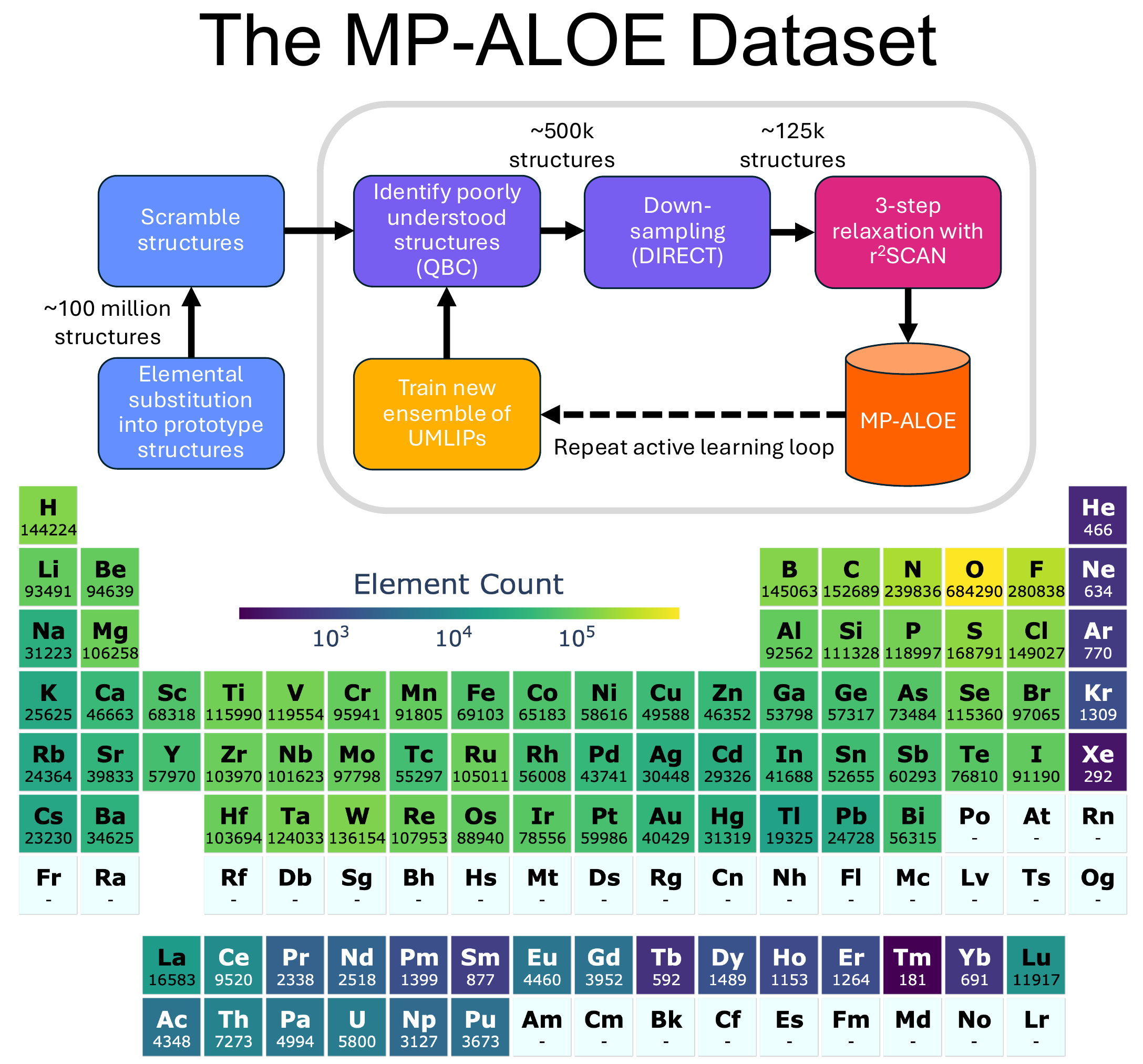}}
	\caption{\textbf{Overview of the \datasetname\, dataset}. (top) The active learning workflow for generating \datasetname. $\sim$100 million structures were generated via elemental substitution into prototype structures. Structures in poorly covered parts of the PES were selected via Query By Committee \cite{seung_query_1992}; these structures were then downsampled using the DIRECT method \cite{qi_robust_2024}. The remaining structures were computed using the r$^2$SCAN functional \cite{furness_accurate_2020}. This process was then repeated. Further details can be found in the Methods section. (bottom) The elemental distribution in \datasetname. Counts are the number of \textit{atoms} of a given element. Visualized using the \texttt{pymatviz} package \cite{riebesell_pymatviz_2022}.}
	\label{fig:summary} 
\end{figure}

We start by detailing the \datasetname\, dataset itself. In total, 909,792 frames of DFT data (from 303,264 structure relaxations) are included in the present work; the workflow for their generation is shown in Fig. \ref{fig:summary} (top). The distribution of atom counts by element for \datasetname\, is shown in Fig. \ref{fig:summary} (bottom); elemental coverage appears to be quite reasonable. 
The over representation of oxygen is consistent with experimental databases like the ICSD \cite{zagorac_recent_2019}, wherein more than half of the experimental inorganic structures contain oxygen.

\begin{figure}[htb]
	\makebox[\textwidth][c]{\includegraphics[width=1.0\linewidth,trim={0mm 0mm 0mm 0mm},clip]{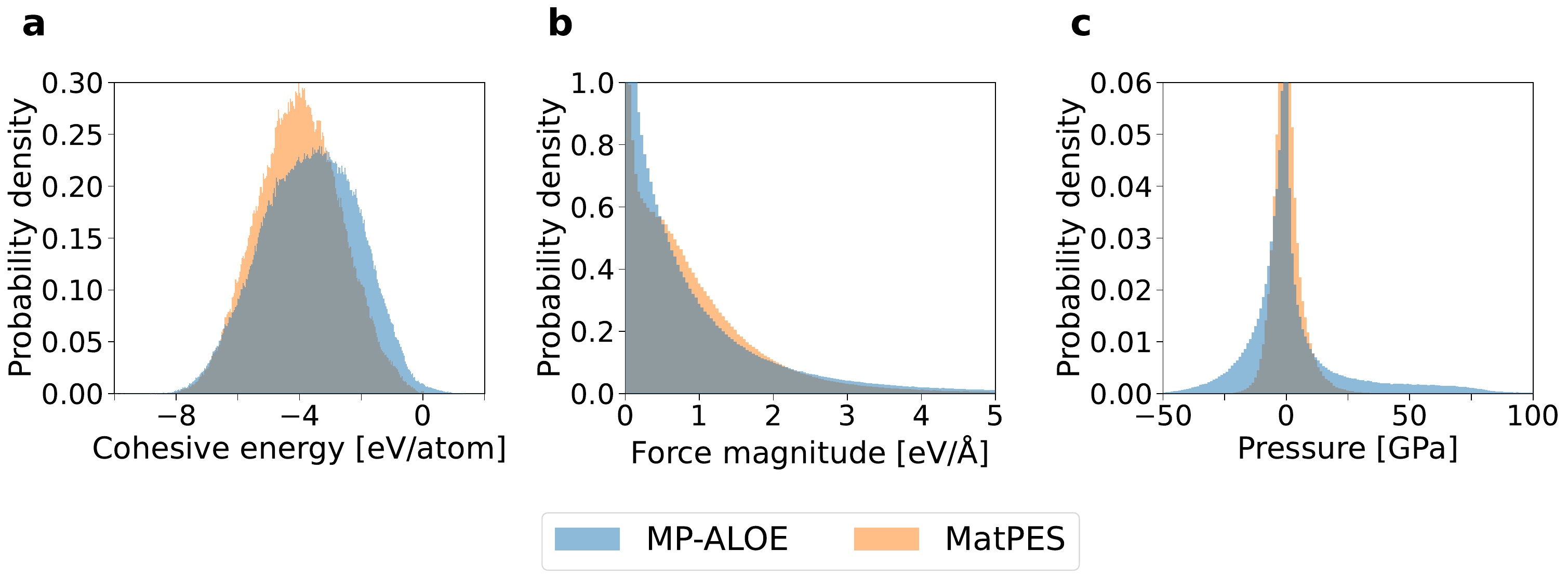}}
	\caption{\textbf{Describing the \datasetname\, dataset}. Distribution of (a) cohesive energies (eV/atom), (b) interatomic force magnitudes (eV/\angstrom), and (c) pressure as derived from one third of the trace of the DFT stress tensor (GPa) in the \datasetname\, and MatPES datasets (the only two public r$^2$SCAN datasets designed for UMLIPs currently available). A total of 3949 points ($\approx$ 0.4\%) in the \datasetname\, dataset have a positive cohesive energy; 3432 of which are explained via the presence of noble gases or the presence of two or more unique nonmetals. For the pressures, positive values correspond to compression.}
	\label{fig:dataset_comparison} 
\end{figure}

The cohesive energies for the \datasetname\, dataset are shown in Fig. \ref{fig:dataset_comparison}a, with the MatPES \cite{kaplan_foundational_2025} dataset included as a reference. \datasetname\, has a higher mean cohesive energy (-3.65 eV/atom) than MatPES (-4.01 eV/atom); \datasetname\, also has a wider distribution of cohesive energies. This was expected, given that MatPES structures are sampled from 300K MD simulations of (largely) near-stable structures from the Materials Project Database \cite{jain_commentary_2013, horton_accelerated_2025}, whereas \datasetname\, was mostly generated combinatorially from unknown hypothetical structures. 

The distribution of forces in \datasetname\, is shown in Fig. \ref{fig:dataset_comparison}b, with MatPES as a reference. Overall, the distributions are roughly comparable, with \datasetname\, having a slightly greater mean force (1.03 eV/\angstrom) than MatPES (0.94 eV/\angstrom). \datasetname\, contains a relatively smooth distribution of forces, and contains a greater sampling of forces over 2 eV/\angstrom\, relative to MatPES.

The distribution of pressures in \datasetname\, is shown in Fig. \ref{fig:dataset_comparison}c, with MatPES as a reference. \datasetname\, has a notably broader spread of pressures, with a notable proportion of pressures between -50 and 100 GPa (as opposed to MatPES, which mostly contains pressures between -20 and 30 GPa).

A discussion of the limitations of the \datasetname\, is included in the Supplemental Information. Additionally, further comparisons of \datasetname\, to the OMat24 dataset \cite{barroso-luque_open_2024} (a notably diverse dataset calculated at the lower PBE level of theory) are included in the Supplemental Information (Fig. S4).

\subsection*{Benchmarking}
\label{subsec:Benchmarking}

Here we compare results between three MACE potentials: a potential trained only on \datasetname, a potential trained only on MatPES, and a potential trained on both of these datasets combined (\datasetname\, + MatPES). Details for how \datasetname\, and MatPES were combined can be found in the Methods section. We use the name of a dataset and the potential trained on that dataset interchangeably to increase readability.

\subsubsection*{Equilibrium Benchmarks}
\label{subsubsec:equilibrium benchmarks} 

Here we compare performance of \datasetname, MatPES, and the combined dataset potentials on predicting equilibrium properties. Specifically, approximately 1000 structures sourced from the WBM dataset \cite{wang_predicting_2021} were relaxed using r$^2$SCAN as part of the MatPES preprint \cite{kaplan_foundational_2025}. The atomic positions of the DFT-relaxed structures were then randomly perturbed by 0.1 \angstrom\, and re-relaxed using the UMLIPs listed; then the cohesive energy error and ``fingerprint distance'' are computed. For the latter, a fingerprint vector is generated by the \texttt{CrystalNN} method \cite{zimmermann_local_2020} for both the DFT-relaxed and UMLIP-relaxed structure. The fingerprint distance is calculated as the euclidean distance between the two fingerprint vectors; a smaller fingerprint distance implies that the DFT-relaxed and UMLIP-relaxed structures are more similar, hence, smaller is better. Further details of the methodology for this benchmark can be found in Ref. \cite{kaplan_foundational_2025}. 

\begin{figure}[htb]
	\makebox[\textwidth][c]{\includegraphics[width=0.5\linewidth,trim={0mm 0mm 0mm 0mm},clip]{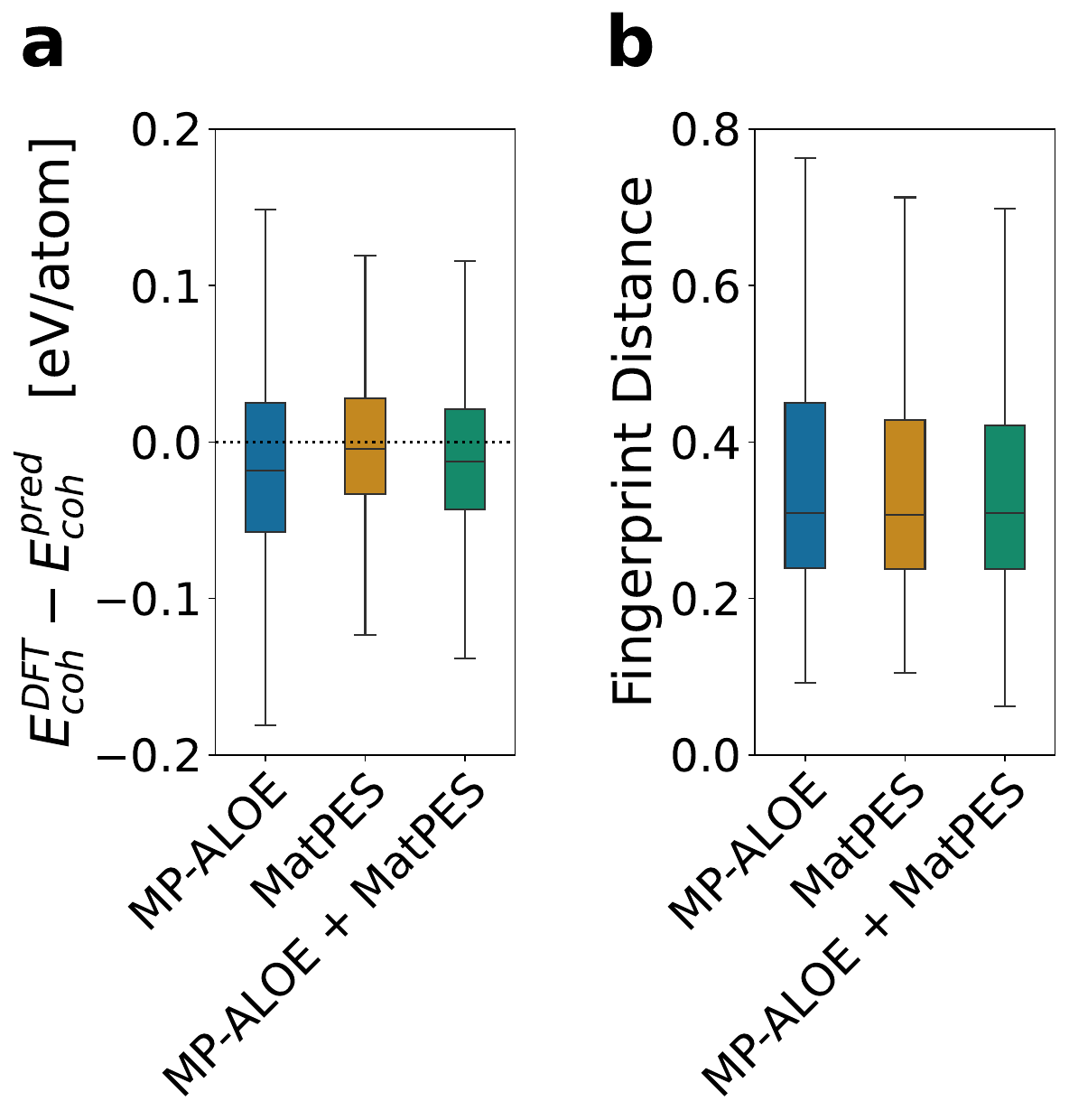}}
	\caption{\textbf{Benchmarking performance of UMLIPs at equilibrium}. (a) Predictions of cohesive energies (eV/atom) and (b) fingerprint distances for equilibrium structures sourced from the WBM dataset \cite{wang_predicting_2021}. Random atomic displacements were applied to DFT-relaxed structures, and then the scrambled structures were re-relaxed using the respective UMLIPs. Fingerprint distance is a measure of structural similarity between the DFT- and UMLIP-relaxed structures; smaller is better. The CrystalNN method used to calculate fingerprint distance is described in Ref. \cite{zimmermann_local_2020}, and further details of this test are described in Ref. \cite{kaplan_foundational_2025}.}
	\label{fig:cohesive and fingerprint} 
\end{figure}

For the cohesive energy task (Fig. \ref{fig:cohesive and fingerprint}a), the MatPES-only model (MAE of 48 meV/atom) performs better than the \datasetname-only model (64 meV/atom); the combined \datasetname\, + MatPES model is quite comparable to the MatPES-only model, with an MAE of 51 meV/atom. This is likely due to the MatPES dataset containing a greater proportion of near-equilibrium data (as shown in Fig. \ref{fig:dataset_comparison}); this is further substantiated by MatPES having a cohesive energy distribution very similar to that of the \textit{explicitly} near-equilibrium MPTrj dataset \cite{deng_chgnet_2023} (see Fig. 2a from Ref. \cite{kaplan_foundational_2025}).

The models are relatively comparable on the fingerprint distance task (Fig. \ref{fig:cohesive and fingerprint}b). This implies that, despite energy predictions showing some differences, the relaxed structures are quite comparable for the three models.

\subsubsection*{Off-Equilibrium Forces}
\label{subsubsec:off-equilibrium forces} 

\begin{figure}[htb]
	\makebox[\textwidth][c]{\includegraphics[width=1.0\linewidth,trim={0mm 0mm 0mm 0mm},clip]{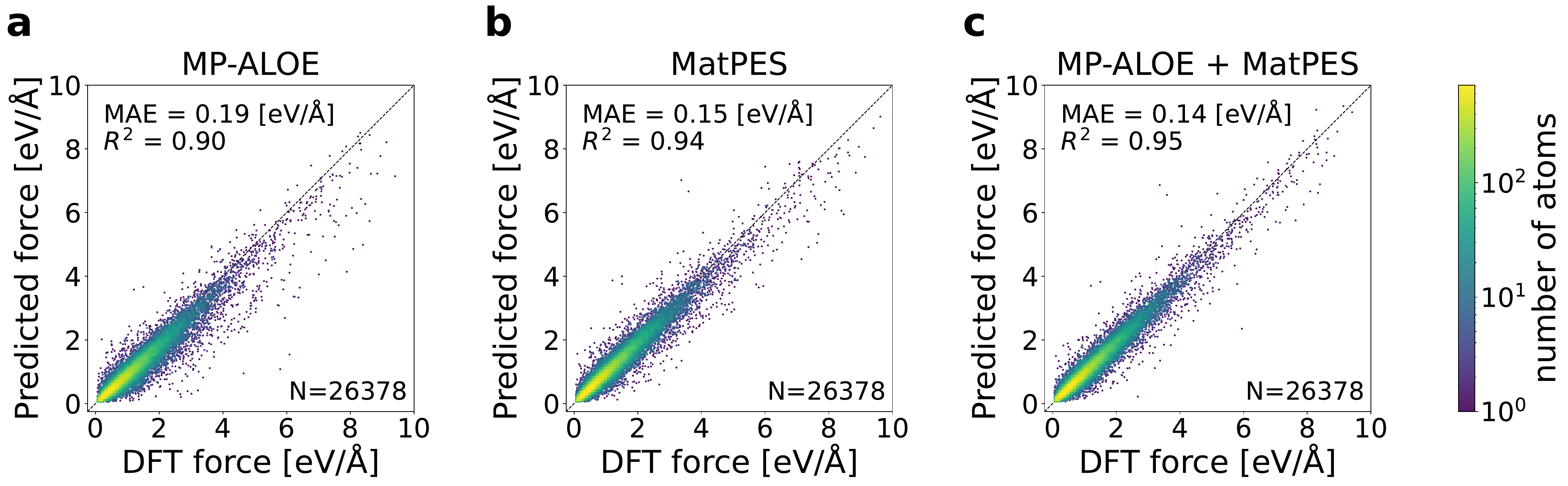}}
	\caption{\textbf{Predictions of off-equilibrium forces (eV/\angstrom)}. Performance of MACE models trained on (a) \datasetname, (b) MatPES, and (c) the combined \datasetname\, + MatPES are shown. The DFT data was generated by taking relaxed structures from the WBM dataset \cite{wang_predicting_2021}, making them into supercells, displacing the atoms by 5\% of the average interatomic distance, and re-computing the forces using the r$^2$SCAN functional.}
	\label{fig:lindemann} 
\end{figure}

UMLIPs have been shown to have poor performance in predicting far-from-equilibrium forces (with a specific trend of underprediction/``softening'')\cite{deng_systematic_2025}. To evaluate this, 1000 structures were sampled from the WBM dataset \cite{wang_predicting_2021}. The selected structures from WBM were then repeated as supercells ranging from 18 to 32 atoms, and the atomic positions of every atom were randomly displaced in a random direction by 5\% of the average interatomic distance; this value was selected because it is half of Lindemann's criterion for melting \cite{lindemann_calculation_1910}.
Predicted vs. actual DFT forces are shown in Fig. \ref{fig:lindemann}. The MatPES-only model performs moderately better than the \datasetname-only model, though both seem to overcome the systematic softening reported on previously \cite{deng_systematic_2025}. Moreover, the combined \datasetname\, + MatPES model achieves marginally higher accuracy than both the MatPES-only and \datasetname-only models.

\subsubsection*{Physicality at Extreme Deformations}
\label{subsubsec:EV_scan_benchmarking}
MLIPs are known to have inconsistent performance at large deformations, sometimes leading to unphysically low energies when far from equilibrium \cite{chiang_mlip_2025}; this is especially prevalent at high compressions. To address this, the energy-volume scan (EV-scan) benchmark from \texttt{MLIP Arena} \cite{chiang_mlip_2025} was developed.
In essence, this task involves a series of \textit{static} calculations of a material that is uniformly strained -- no relaxation of the cell nor of the atom positions is permitted. Given that no relaxation is allowed, one can expect a structure's energy to monotonically increase as it is deformed away from its equilibrium configuration. 
Indeed, first principles evidence for this behavior at extreme strain can be found in, e.g., Refs. \citenum{alchagirov_sjeos_2001,kaplan_ksturn_2021}.
The EV-scan benchmark quantifies this behavior for a set of 1000 structures from the WBM dataset \cite{wang_predicting_2021} when subjected to extreme deformations of $\pm$20\% along each lattice direction (all lattice vectors are scaled uniformly). When all three lattice vectors are scaled by the same numeric factor, the energy is expected to have a single local extremum (minimum), hence the derivative should change sign exactly once.
This is distinct from the case of, e.g., Bain paths, whereby scaling the lattice vectors by unequal values can produce multiple extrema on the potential energy surface \cite{grimvall_metals_2012}.
For the MLIP benchmark, cases where the derivative changes sign more than once are counted as failures. Moreover, the energy should monotonically increase as a material is deformed away from equilibrium, corresponding to a Spearman's coefficient of -1 for compressive strain. $\frac{\partial E}{\partial V}$ should monotonically increase with application of expansive strain, corresponding to a Spearman's coefficient of +1.

Results for MACE potentials trained on \datasetname, MatPES, and both datasets combined are presented in Table \ref{tab:EV_scan}. The original MACE-MP-0a \cite{batatia_foundation_2023} is also included as a reference point. Overall, the model trained only on \datasetname\, performs notably better than the model trained only on MatPES. However, the combined dataset performs the best on all categories.

\begin{table}[htbp]
    \fontsize{12pt}{12}\selectfont
    \centering
    \caption{\textbf{Physicality of the potential energy surface under extreme hydrostatic strain.} Performance on the Energy-Volume scan benchmark from \texttt{MLIP Arena}, which evaluates structures undergoing a series of deformations of $\pm$ 20\% in each lattice direction, using only single point calculations at each point. There should be a singular global minimum, as the atom positions and lattice are held fixed; the ``Failures percentage'' metric calculates the percentage of structures where the derivative changes sign more than once. Ideal performance on the ``$E$ under compression'' metric is -1, as energy should monotonically increase as a material is compressed. Ideal performance on the ``$\frac{\partial E}{\partial V}$ under compression'' metric is +1, as $\frac{\partial E}{\partial V}$ should go from negative values, to zero at the equilibrium volume, to positive values as the structure is dissociated to isolated atoms. We once again emphasize that these $E-V$ scans do \textit{not} involve relaxation, hence no structural transformations occur during these tests.}
    \label{tab:EV_scan}
    \begin{tabular}{lccc}
    \toprule
    \multirow{2}{*}{Model} &  Failures & \multicolumn{2}{c}{Spearman's coefficient$^a$} \\
    & percentage $\downarrow$ & $E$: compression $\downarrow$ & $\frac{\partial E}{\partial V}$: compression $\uparrow$ \\
    \cmidrule(r){1-1}\cmidrule(lr){2-2}\cmidrule(lr){3-4}
    MACE-MP-0a           & 10.6\%          & -0.944          & 0.901 \\
    \datasetname           & 2.5\%          & -0.990          & 0.980 \\
    MatPES                 & 14.8\%          & -0.919          & 0.857 \\
    \datasetname\, + MatPES & \textbf{0.8\%} & \textbf{-0.995} & \textbf{0.994} \\
    \bottomrule
    \end{tabular} 
    \\
    \footnotesize{\hspace{5mm} $^a$ values are bounded by [-1,1].}
\end{table}

\subsubsection*{Molecular Dynamics Stability}
\label{subsubsec:MD stability} 

A desirable property of UMLIPs is their relative computational affordability to perform longer molecular dynamics (MD) simulations, and of larger simulation cells. The MD stability benchmark task from \texttt{MLIP Arena} \cite{chiang_mlip_2025} attempts to measure this by again subjecting a reasonable structure to extreme environmental conditions. The structures used for this benchmark are taken from the RM24 structures in Ref. \cite{chiang_mlip_2025}, which were generated as random mixtures of stable materials from the Materials Project using \texttt{PackMol} \cite{martinez_packmol_2009}.
To obtain reasonable initial structures, they were then relaxed using a Ziegler, Biersack, Littmark (ZBL) potential \cite{ziegler_biersack_littmark_1983,ziegler_stopping_1985}. For the benchmark, a UMLIP is used to perform MD simulations under increasingly extreme environments. In total, 100 structures (with an average of 5-6 elements and 525 atoms) are simulated for each of the following two sub-tasks.
Further details on these benchmark tasks can be found in Ref. \cite{chiang_mlip_2025}.

\begin{figure}[htb]
	\makebox[\textwidth][c]{\includegraphics[width=1.0\linewidth,trim={0mm 0mm 0mm 0mm},clip]{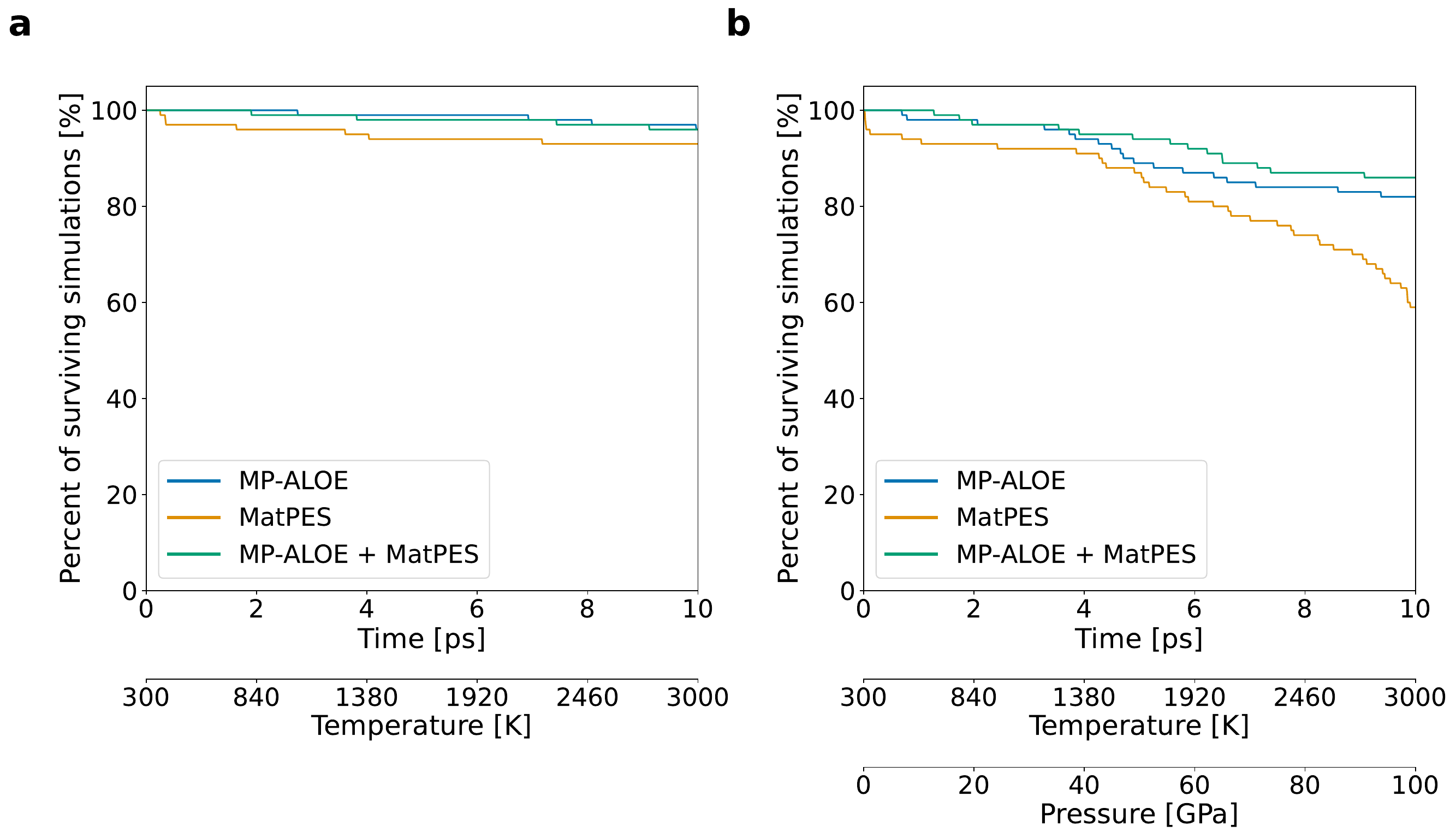}}
	\caption{\textbf{Stability during molecular dynamics simulations}. (a) 100 NVT simulations with a linear temperature ramp from 300 to 3000 K. (b) 100 NPT simulations with a linear temperature ramp from 300 to 3000 K \textit{and} a linear pressure ramp from 0 to 100 GPa. Tasks and the randomly-mixed structures were sourced from \texttt{MLIP Arena} \cite{chiang_mlip_2025}. The ordinate displays the percentage of surviving runs with a ``valid'' MD trajectory at a given timestep. 
    See the text for a description of ``valid''.
    }
	\label{fig:MD stability} 
\end{figure}

The first sub-task involves NVT simulations with linear temperature scaling from 300 to 3000 K over 10 ps, with results shown in Fig. \ref{fig:MD stability}a. All three datasets perform relatively well on the NVT sub-task. \datasetname\, performs the best with 98.8\% of scheduled timesteps completed, compared to MatPES with 94.7\% and the combined \datasetname\, + MatPES with 98.2\% completion.
Here, a given trajectory frame is deemed ``valid'' simply if the energy is within 100 eV/atom of the initial frame, the maximum kinetic energy for any atom is less than 100 eV.
For the NPT simulations, we also demand that the cell volume stay bounded within one-tenth of and ten times the initial volume to be considered valid.
These constraints help ensure that a simulation does not veer into a ballistic regime, nor that the structures dissociate (explode).

The second sub-task involves NPT simulations with the same 300 to 3000 K temperature scaling \textit{and} linear pressure scaling from 0 to 100 GPa over 10 ps. In Fig. \ref{fig:MD stability}b, the \datasetname-only model significantly outperforms the MatPES-only model, completing 90.6\% and 83.7\% of scheduled timesteps, respectively. This can likely be attributed to the greater sampling of larger pressures in \datasetname, as shown in Fig. \ref{fig:dataset_comparison}c. The combined \datasetname\, + MatPES-trained model further improves on this, completing 93.2\% of scheduled timesteps.

Note that we modify the original NPT benchmark from \texttt{MLIP Arena} to reduce the maximum pressure to 100 GPa instead of 500 GPa to better reflect physically attainable pressures. Results for the original benchmark are shown in Supplemental Fig. S5, which demonstrates that the \datasetname-only model successfully completes twice as many MD timesteps as the MatPES-only model when the pressure is increased to 500 GPa.

\section*{Discussion}
\label{sec:Discussion}

We present a new dataset for universal machine learning interatomic potentials (UMLIPs) trained on the accurate r$^2$SCAN functional \cite{furness_accurate_2020}.
The dataset was constructed via a form of active learning (query by committee \cite{seung_query_1992}) on primarily hypothetical compositions, thus we call it \datasetname\, (\textbf{M}aterials \textbf{P}roject - \textbf{A}ctive \textbf{L}earning of \textbf{O}ff \textbf{E}quilibrium structures). We compare the presented \datasetname\, to MatPES \cite{kaplan_foundational_2025} (the only other public r$^2$SCAN dataset designed specifically for UMLIPs at the time of writing).
By construction, both datasets were calculated using directly compatible plane-wave density functional theory (DFT) calculation parameters. 
To expand the space of possible but realistic chemical environments sampled in the dataset, we performed elemental substitution on a set of prototype structures, emphasizing bonding elements and non-$f$-block elements (which are likely to be poorly described by frozen core pseudopotentials and display less diverse chemical arrangements).

We then trained three MACE potentials \cite{batatia_mace_2022} on \datasetname, MatPES, and the combined \datasetname\, + MatPES datasets. We compared their performance in predicting the cohesive energies of equilibrium structures and the magnitudes of off-equilibrium forces.
For these two tasks, all three models perform comparably, however the MatPES-only model tends to predict slightly more accurate cohesive energies, and the combined \datasetname + MatPES model clearly achieves the best performance overall on these three tasks.
Importantly, we note that the \datasetname- and MatPES-only models perform almost identically in equilibrating structures to a reference configuration, indicating that they sample similar ranges of near-equilibrium interatomic forces.

The third benchmark task examined the physicality of the shapes of the potentials' energy-volume curves at extreme hydrostatic strain. Here, \datasetname\, significantly outperforms MatPES, with 2.5\% and 15\% of $E-V$ curves failing to meeting physicality criteria, respectively. Moreover, the combined \datasetname\, + MatPES shows the best performance overall across all categories.

The final benchmark task demonstrates the true value of \datasetname\, via stability of molecular dynamics (MD) calculations under extreme ensemble conditions.
In both NVT runs with the temperature increasing from 300 to 3000 K, and NPT runs with simultaneously increasing temperature and pressure from 0 to 100 GPa, the \datasetname-only model significantly outperforms the MatPES-only model.
Once more, the combined \datasetname\, + MatPES model demonstrates the highest stability in extreme environments.

In the nascent space of UMLIP research, more systematic methods to improve the quality of the datasets used to train them is highly needed.
We have established a systematic and computationally tractable way to perform active learning-reinforced sampling, thereby targeting unexplored sections of the potential energy surface at a tractable computational cost.
Our publicly-available \datasetname\, dataset 
is fully compatible with MatPES, and is recommended for use in training UMLIPs.

\section*{Methods} \label{sec:Methods}

The overarching flow for generating the \datasetname\, dataset is as follows.

\subsection*{Generating Structures}
\label{subsec:generating structures}

Most of the structures contained in this dataset were created via the unbiased substitution of elements into prototype structures. Note that, as the probability of sampling an element was uniform, structures that are not conventionally charge balanced could be generated; this is explored and mostly confirmed in Supplemental Table S1. Prototypes contained within the ICSD \cite{zagorac_recent_2019} and Materials Project \cite{jain_commentary_2013, horton_accelerated_2025} were chosen, wherein the \texttt{StructureMatcher} from \texttt{pymatgen} \cite{ong_python_2013} was used to remove duplicates. To ensure computational efficiency during the later DFT calculations, the prototypes considered were restricted to be between 2-8 atoms. Moreover, to reduce combinatorial explosion when occupying the prototypes with all combinations of elements included in this dataset, only up to ternary structures were considered. The result is 817 prototype structures: 660 binaries and 157 ternaries. When all possible substitutions of the 89 elements contained in MP are enumerated, this totals over 100 million structures to sample from. 

The initial lattice parameters for these structures were estimated using the method described in the Supplemental Information. Then, to obtain a broader distribution of forces and stresses, the atomic positions were randomly displaced and the lattice vectors were randomly scaled according to the procedure in Supplemental Information.
Supplemental Table S1 demonstrates that this procedure reliably produces structures with minimum nearest neighbor distances greater than the overlap region between pseudopotential cores.

\subsection*{Query By Committee}
\label{subsec:query by committee}

Query By Committee (QBC) \cite{seung_query_1992} is an established active learning technique that has been applied across a wide range of fields, including interatomic potentials (e.g. \cite{smith_less_2018}). Here, we apply the technique to select structures that cover unexplored portions of the PES. Our application of the QBC technique is described as follows.

The generated structures from the previous subsection are fed into an ensemble of interatomic potentials which predict their properties. If the ensemble `disagrees' above some heuristically set threshold on the predicted energy, forces, \textit{and/or} stress, the structure is selected for the down-sampling step (detailed in the next subsection). 
The criterion for disagreement used was when the standard deviation of the ensemble's predicted energies ($\sigma_{e}$), forces ($\sigma_{f}$), or stresses ($\sigma_{s}$) are greater than 100 meV/atom, 100 meV/\r{A}, or 100 meV/\r{A}$^3$, respectively; these values roughly correspond to the errors of MACE-MP-0 on its own test set. To be explicit, the ensemble of models must disagree on \textit{at least} one of the energies, forces, or stresses for a structure to be selected.
To ensure that elements which typically do not bond (i.e., the noble gases and technetium) and elements which are unlikely to be well-described by frozen-core pseudopotentials (i.e., the $f$-block) were not over-represented in the dataset, a heuristic factor was applied to increase the threshold for committee consensus.

For the first iteration of active learning, existing interatomic potentials were used. Namely, the ensemble was comprised of the pre-trained MACE-MP-0 \cite{batatia_foundation_2023}, CHGNet \cite{deng_chgnet_2023}, and M3GNet \cite{chen_universal_2022} models. 
Both to reduce the computational burden of model training and because no existing r$^2$SCAN UMLIP was available when the first active learning cycle was performed, models trained to PBE data were used to select structures for the solely-r$^2$SCAN-based \datasetname\, dataset. These particular three pretrained models were selected because they were the top three performers on the Matbench Discovery benchmark for UMLIPs \cite{riebesell_matbench_2024} at the time this project was started. For the remaining iterations of active learning, three MACE models were trained (details for training can be found later in the Methods section). 
MACE was chosen primarily because it maintains a higher body-order than M3GNet or CHGNet.

This methodology resulted in roughly $\sim$500,000 structures being selected in a given active learning cycle; these 500,000 structures were then downsampled in the next subsection.

\subsection*{Downsampling}
\label{subsec:downsampling}

One noteworthy drawback of batch-mode active learning is that the batch of data selected for labelling often has informational overlap \cite{ren_survey_2022}. To mitigate this, a method for downsampling a diverse subset (from the larger set of identified structures from the QBC) is needed. DIRECT sampling \cite{qi_robust_2024} was identified as a suitable method for downsampling, as it has been successfully demonstrated to select diverse samples that can create comparable models with substantially less data. The DIRECT method involves the following steps: 1) all structures are featurized using the M3GNet Formation Energy model \cite{chen_universal_2022} (accessed via the \texttt{matgl} repository \cite{ko_materials_2025}); 2) principal component analysis is then performed for dimensionality reduction; 3) the dimensionally reduced features are then clustered using BIRCH clustering \cite{zhang_birch_1996}; and 4) a fixed number of structures are selected from each cluster. For a given active learning cycle, the use of DIRECT sampling reduced the number of selected samples from roughly $\sim$500,000 down to roughly $\sim$125,000 structures. These $\sim$125,000 structures are then simulated using DFT, as described in later in the Methods section. 

\subsection*{Additional Data}
\label{subsec:Additional Data}

The structures generated via the process described above are often far from equilibrium. To augment this aspect of the data generation process, a small amount of additional near-equilibrium data from the Materials Project was also included. Specifically, all structures with up to 3 elements and up to 32 atoms (totalling roughly $\sim$30,000 structures) were re-calculated using the same DFT settings as those used to calculate the structures selected via QBC (described in the next subsection).

\subsection*{Density Functional Theory Calculation Details}
\label{subsec:density functional theory calculation details}

DFT calculations were performed using the Vienna Ab-Initio Simulation Package (VASP) \cite{kresse_ultrasoft_1999} using projector-augmented wave (PAW) potentials \cite{blochl_projector_1994, kresse_efficient_1996}. A plane-wave cutoff energy of 680 eV was used, with the KSPACING parameter set to 0.2. Further input parameters used are detailed in the MP24RelaxSet from \texttt{pymatgen}. 

Calculations were performed in two stages. First, a static calculation using the PBE exchange-correlation functional \cite{perdew_generalized_1996} was performed. Then, the WAVECAR from the static PBE calculation is fed into a relaxation calculation performed using the r$^2$SCAN functional \cite{furness_accurate_2020}. The second r$^2$SCAN calculation always ran for \textit{three} ionic steps; this was chosen so that initial structures (which were often relatively far from equilibrium) have a chance to equilibriate to a reasonable pressure, less extreme forces, etc. This is especially necessary given the difficulty of estimating the lattice parameter of an unknown material. The authors chose to perform three ionic steps intentionally---this was the number of ionic steps required for 90\% of calculations to include both a positive and negative pressure in at least one ionic step for a set of test calculations. 

In total, the above-prescribed workflow converged for 82\% of structures. All calculations and workflows were performed using the \texttt{atomate2} package \cite{ganose_atomate2_2025}, which itself builds upon the \texttt{fireworks} \cite{jain_fireworks_2015}, \texttt{jobflow} \cite{rosen_jobflow_2024}, and \texttt{pymatgen} \cite{ong_python_2013} packages.

\subsection*{Model Training}
\label{subsec:model training}

After each iteration of active learning, the r$^2$SCAN data generated was compiled and used to train three MACE potentials \cite{batatia_mace_2022}. All input parameters for the models trained in this work were (nearly) identical to those used in the `large' model from the original version of the MACE-MP-0 manuscript on arXiv \cite{batatia_foundation_2023} (now referred to as MACE-MP-0a, which was released with v0.3.6 of the MACE package on github). The only parameters changed were the isolated atom energies, which were recomputed using r$^2$SCAN. Each potential had exactly 5,725,072 parameters. A 90-5-5 train-validation-test split was used, and the models were trained for 100 epochs (which was empirically selected based on when learning appeared to plateau). The only noteworthy difference between the three models within a given active learning iteration was that the training/validation/test splits were randomized, thus creating small variations between the models.

\subsection*{Combining \datasetname\, and MatPES}
\label{subsec:combining datasets}

\datasetname\, and MatPES were calculated using directly compatible DFT parameters. However, both datasets had a small proportion of structures sourced from the Materials Project included, with some MP structures appearing in both datasets. Hence, the datasets cannot simply be merged, as this would lead to data redundancy. MatPES, by construction, contains a greater number of MP-sourced structures, including all of the MP-sourced structures in \datasetname. Hence, only the \textit{non-MP} structures in \datasetname\, were merged with \textit{all} r$^2$SCAN calculations within MatPES during training of the ``\datasetname\, + MatPES'' potential.

\section*{Acknowledgements}
\label{sec:Acknowledgements}

This work was intellectually led by the Materials Project program KC23MP, supported by the U.S. Department of Energy, Office of Science, Office of Basic Energy Sciences, Materials Sciences and Engineering Division under contract No. DE-AC02-05-CH11231.

Matthew Kuner was supported in part by the National Science Foundation Graduate Research Fellowship Program under Grant No. DGE-2146752. Any opinions, findings, and conclusions or recommendations expressed in this material are those of the author(s) and do not necessarily reflect the views of the National Science Foundation.

This research used the Savio computational cluster resource provided by the Berkeley Research Computing program at the University of California, Berkeley (supported by the UC Berkeley Chancellor, Vice Chancellor for Research, and Chief Information Officer). 

This research used the Lawrencium computational cluster resource provided by the IT Division at the Lawrence Berkeley National Laboratory (Supported by the Director, Office of Science, Office of Basic Energy Sciences, of the U.S. Department of Energy under Contract No. DE-AC02-05CH11231)

This research used resources of the National Energy Research Scientific Computing Center (NERSC), a U.S. Department of Energy Office of Science User Facility located at Lawrence Berkeley National Laboratory, operated under Contract No. DE-AC02-05CH11231 using NERSC award BES-ERCAP-0022838.

The authors appreciate the insightful discussions from Mr. Yuan Chiang, Dr. Shivani Srivastava, Professor Bingqing Cheng, and Professor Mary Scott at UC Berkeley, and from Dr. Anubhav Jain at Lawrence Berkeley National Laboratory.

\section*{Author Contribution Statement}
\label{sec:author contribution statement}

M.C.K.: conceived of the project, performed the DFT calculations, trained the UMLIPs, performed the benchmark calculations, created visualizations, wrote and edited the manuscript. A.D.K.: contributed to benchmark conception, created visualizations, assessed the validity of the DFT calculations, wrote and edited the manuscript. K.A.P.: manuscript editing, provision of computational resources. M.A.: project design, manuscript editing, provision of computational resources. D.C.C.: project design, manuscript editing, provision of computational resources.


\section*{Declaration of Competing Interest}
\label{sec:competing interests}
The authors declare that they have no known competing financial interests or personal relationships that could have appeared to influence the work reported in this paper.

\section*{Data Availability}
\label{sec:data availability}
The \datasetname\, dataset can be downloaded at \url{https://doi.org/10.6084/m9.figshare.29452190}. The raw VASP outputs, totalling roughly 50 TB, can be made available upon reasonable request; WAVECAR files were \textit{not} kept. The trained MACE potentials are also available at the URL above, as is the data for the off-equilibrium forces benchmark. All other materials used in the creation of this work are available upon reasonable request.


\clearpage

\bibliography{references}
\bibliographystyle{naturemag}

\clearpage










\renewcommand{\thepage}{S\arabic{page}}
\renewcommand{\thesection}{S\arabic{section}}
\renewcommand{\theequation}{S\arabic{equation}}
\renewcommand{\thetable}{S\arabic{table}}
\renewcommand{\thefigure}{S\arabic{figure}}








\setcounter{figure}{0}
\setcounter{table}{0}
\setcounter{page}{1}

\section*{Supplemental Information: \papername}

\subsection*{Dataset Limitations}
\label{sec:Dataset Limitations}

The presented \datasetname\, dataset is the largest r$^2$SCAN dataset published to date, and shows an improvement in diversity of the sampled energies, forces, and pressures over previously published datasets. However there is always room for growth. \datasetname\, primarily contains small bulk crystals (mostly 2-8 atoms, plus a small sample of structures from the Materials Project), foregoing the larger simulation cells needed to capture point defects, dislocations, grain boundaries, surfaces, non-crystalline materials, etc.
For simulating such chemical environments, it is likely that end users would benefit from fine tuning an \datasetname-trained model with bespoke data relevant for their use case.

Further, while the level of DFT approximation, r$^2$SCAN, used in \datasetname\, and MatPES is generally the highest level which is tractable in high-throughput solid-state calculations, there are still higher levels of approximation which would benefit MLIPs.
It is expected that hybrids would outperform meta-GGAs in predicting certain solid-state properties accessible to UMLIPs, such as defect formation energies \cite{maciaszek_nv_2023}.
They also make improvements in electronic structure which cannot be accessed by current UMLIPs.
However, their computational complexity in plane-wave codes and non-systematic improvement over meta-GGAs limits their practical use in solid-state calculations.
This is in complete contrast to molecular thermochemistry, where localized basis sets permit faster evaluation of exact exchange needed for hybrids, and where hybrids markedly improve over meta-GGAs for molecular thermochemistry \cite{goerigk_gmktn_2017}.

By construction, \datasetname\, also under-emphasizes the prevalence of $f$-block elements.
It has been established that frozen-core pseudopotentials struggle to realize the diverse bonding arrangements possible for this group of elements \cite{bosoni_pseudo_2023}, even with relativistic effects included for the pseudocore.
These materials also adopt many near-degenerate spin phases which cannot currently be captured by UMLIPs, see e.g., Ref. \cite{lane_iridate_2020} and references therein.
However, future improvements to the dataset might include expansion of the $f$-block chemistries included.

\subsection*{Guessing Lattice Parmeters}
\label{sec:guessing lattice parameters}

Accurately estimating the lattice parameter of a material \textit{a priori} without performing a DFT-level calculation is a yet-unsolved problem. The authors note that further improved UMLIPs may be one promising solution to this problem, but the current state-of-the-art UMLIPs are not yet capable of such tasks reliably. The procedure for estimating the lattice constants is as follows:

First, the neighbors of all atoms within a sphere of roughly 8 \r{A} was examined; the smallest distance between each pair of elements is found (e.g. for a binary A-B material, the smallest A-A distance, smallest A-B distance, and the smallest B-B distance were found). Then, a set of estimated interatomic distances was calculated for each pair of elements in the material. The material's volume was then scaled such that no bonds were below the relevant estimated atomic radii. Several sets of atomic radii were tested, and the \texttt{pymatgen.element.atomic\_radius} was selected (with minor modifications) as it tended to produce structures closest in final geometries to the DFT-relaxed structures.
The ionic radii, which we did \textit{not} select for this purpose, produced structures with systematically larger volumes than in their relaxed configurations.

\begin{figure}[htb]
	\makebox[\textwidth][c]{\includegraphics[width=0.6\linewidth,trim={0mm 0mm 0mm 0mm},clip]{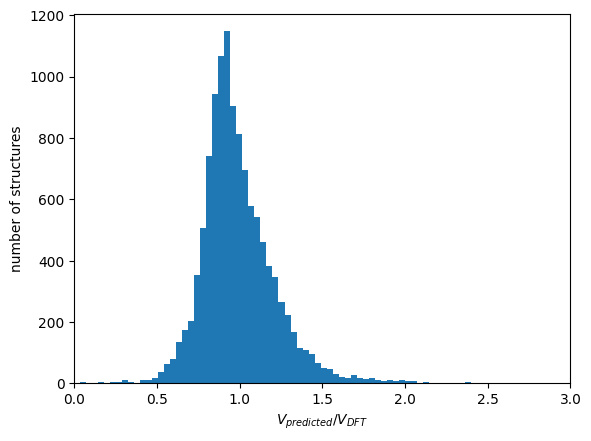}}
	\caption{Ratio of predicted- to DFT-volumes for 11,655 materials from MP.}
	\label{fig:guessing_lattice_params} 
\end{figure}

This method was tested on roughly 10,000 materials from Materials Project; the distribution of $V_{predicted} / V_{DFT}$ is centered at 1 with a standard deviation of roughly 0.2, as shown in Fig. \ref{fig:guessing_lattice_params}. There were no noteworthy systematic errors associated with over- or under-estimating the volumes of particular elements.

The authors acknowledge that our prescribed lattice parameter estimation method is similar to the method presented by Chu \textit{et al.} \cite{chu_predicting_2018}; unfortunately, this was brought to our attention after the dataset had been completed.

\clearpage
\subsection*{Scrambling Structures}
\label{sec:scrambling structures}

To ensure a diverse range of energies, forces, and stresses, the structures were randomly distorted and the atoms were randomly translated from their ideal positions in the prototype. Specifically, the atoms had their atomic positions scrambled along each lattice vector using values drawn from a normal distribution with standard deviation of 2\% of the lattice vector length; this was selected so that the median force in our dataset was approximately 1 eV/\r{A}. Moreover, the lattice vectors then underwent deformation along all 6 strain modes, each with strain magnitudes drawn from a normal distribution with a standard deviation of 2\%; this was selected to increase the diversity of stresses calculated.





\subsection*{Further exploration of the \datasetname\, dataset}
\label{sec:further exploration of the dataset}

\begin{figure}[htb]
	\makebox[\textwidth][c]{\includegraphics[width=01\linewidth,trim={0mm 0mm 0mm 0mm},clip]{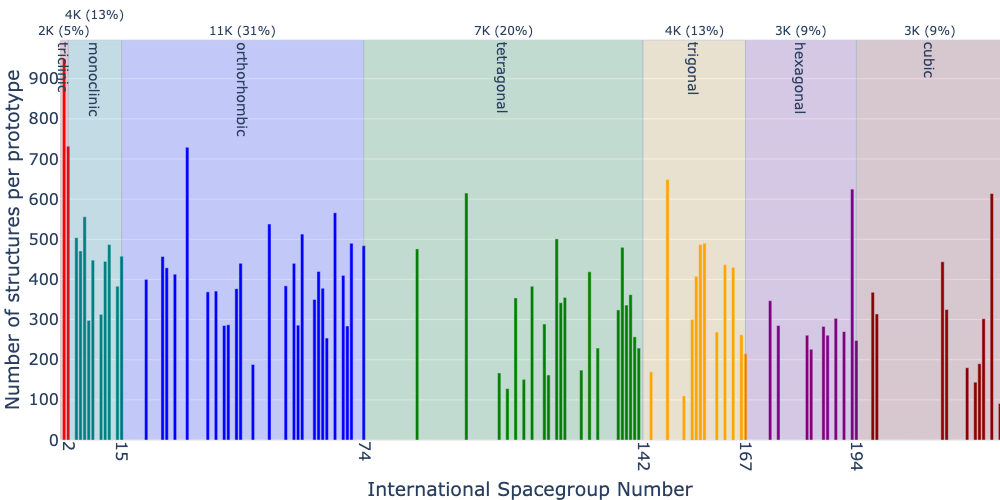}}
	\caption{Distribution of space groups for \emph{initial} structures with successful DFT calculations in \datasetname. Note that we are plotting the space group of the \textit{underlying prototype} before the structure is scrambled.}
	\label{fig:spacegroup_distribution} 
\end{figure}

The distribution of the pre-scramble structures' space groups in \datasetname\, is shown in Fig. \ref{fig:spacegroup_distribution}. The number of structures per space group is approximately proportional to the number of prototypes from that space group, implying a relatively high degree of structural diversity. Note that all of the non-Materials Project structures in \datasetname\, (which is roughly 90\% of \datasetname) are triclinic; here, we are plotting the space group of the \textit{underlying prototype} before the structure is scrambled.

\begin{figure}[htb]
	\makebox[\textwidth][c]{\includegraphics[width=0.6\linewidth,trim={0mm 0mm 0mm 0mm},clip]{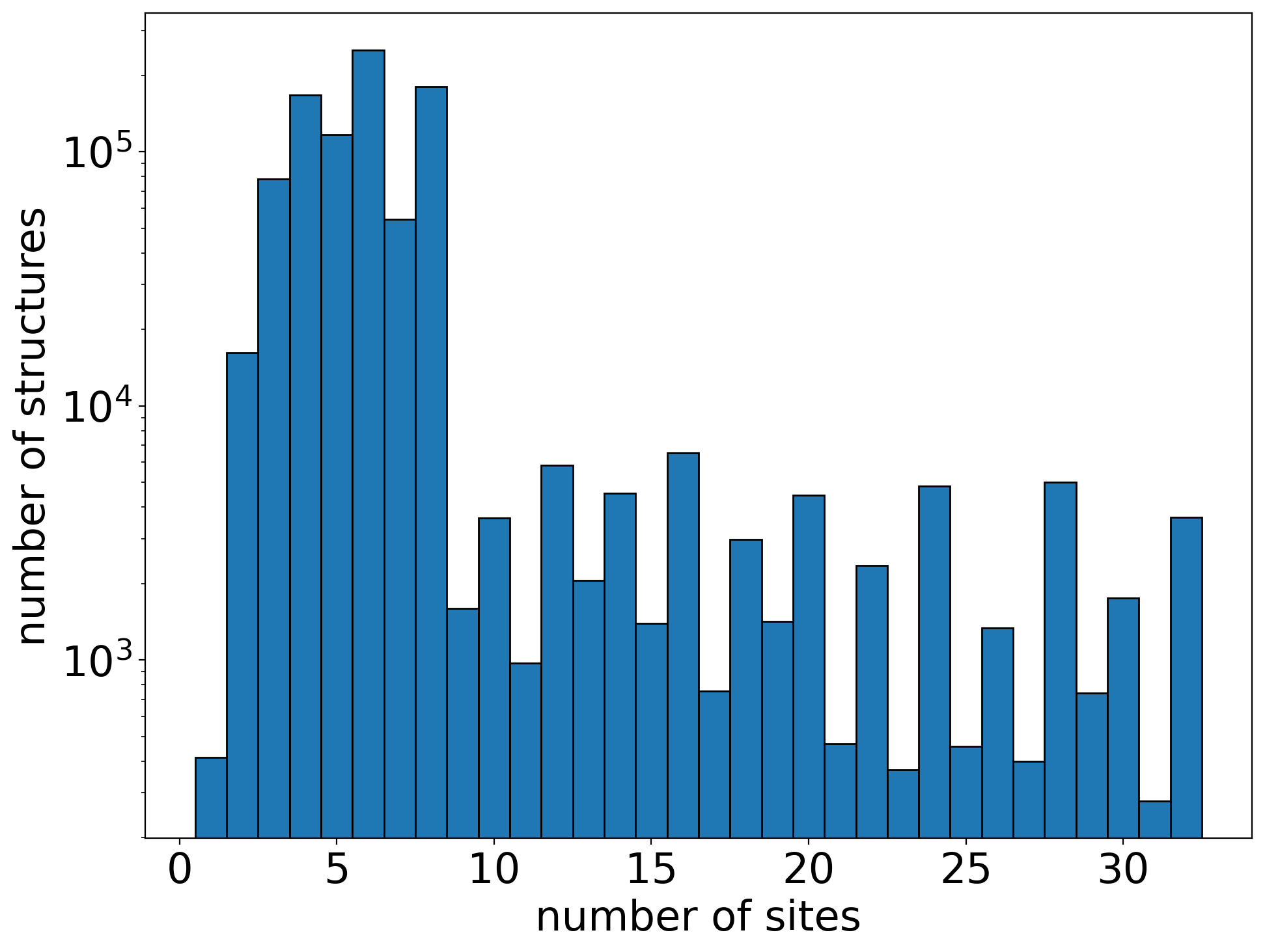}}
	\caption{Distribution of number of atomic sites per structure for structures with successful DFT calculations in \datasetname.}
	\label{fig:number_of_sites_distribution} 
\end{figure}

The distribution of the number of atomic sites per structure is shown in Fig. \ref{fig:number_of_sites_distribution}. As prescribed, the vast majority of structures have between 2--8 atoms, with the remaining structures being the added equilibrium MP-sourced structures.

\clearpage
\subsection*{Further comparison of the \datasetname\, dataset to other notable datasets}
\label{sec:further comparison of the dataset to competitors}

\begin{figure}[htb]
	\makebox[\textwidth][c]{\includegraphics[width=1.0\linewidth,trim={0mm 0mm 0mm 0mm},clip]{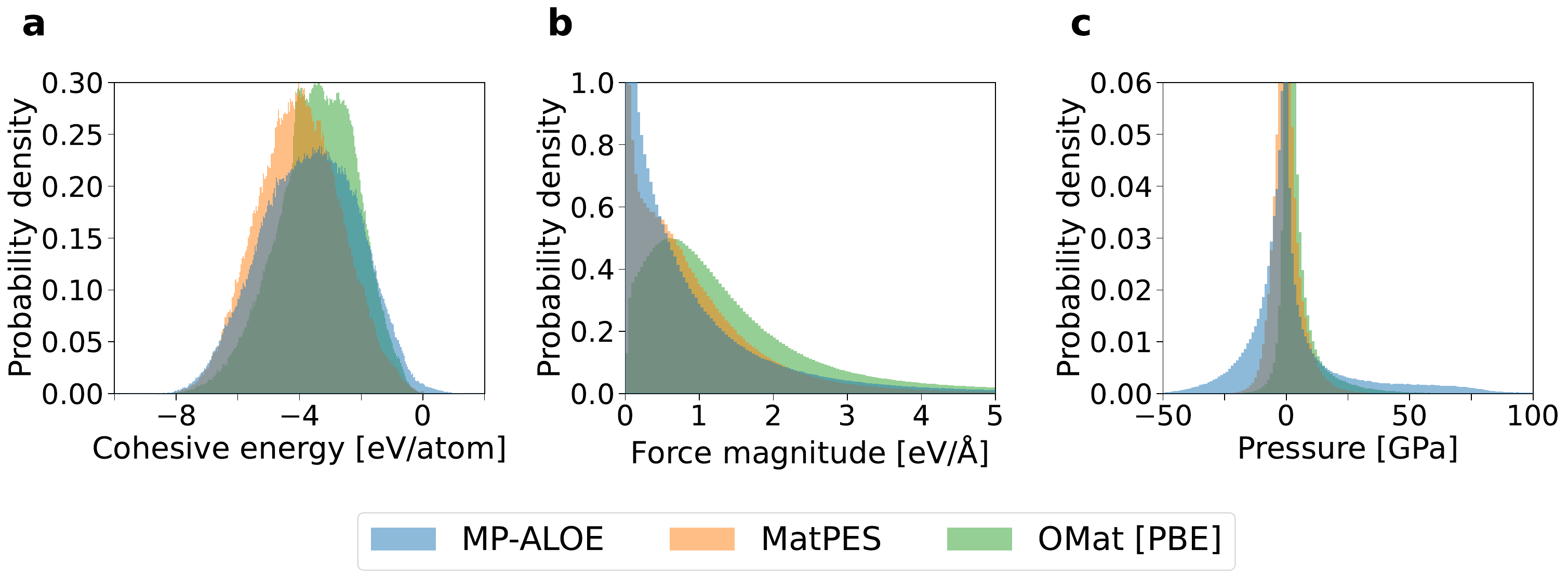}}
	\caption{\textbf{Describing the \datasetname\, dataset (with additional comparisons not in the main text)}. Distribution of (a) cohesive energies (eV/atom), (b) interatomic force magnitudes (eV/\angstrom), and (c) pressure as derived from one third of the trace of the DFT stress tensor (GPa) in the \datasetname\, and MatPES datasets (the only two open-source r$^2$SCAN datasets designed for UMLIPs currently available). The OMat dataset is also included here to provide additional context, despite OMat being computed at the less accurate PBE level of theory. A total of 3949 points ($\approx$ 0.4\%) in the \datasetname\, dataset have a positive cohesive energy; 3432 of which are explained via the presence of noble gases or the presence of two or more unique nonmetals. For the pressures, positive values correspond to compression.}
	\label{fig:dataset_comparison_including_OMat} 
\end{figure}

The \datasetname\, is compared to MatPES \textit{and} OMat \cite{barroso-luque_open_2024} in Fig. \ref{fig:dataset_comparison_including_OMat}. This figure is analagous to Fig. 2 from the main text, with the inclusion of OMat for additional context; we only plot the \textit{validation} set from OMat. Note that OMat is not compatible with \datasetname\, or MatPES, as OMat computed with the PBE GGA and a different set of pseudopotentials. 

In Fig. \ref{fig:dataset_comparison_including_OMat}a, \datasetname\, demonstrates a wider diversity of cohesive energies than OMat, including notably higher proportions of both high and low energy structures. To describe the cohesive energy distributions quantitatively, the \datasetname\,, MatPES, and OMat datasets have a mean and standard deviation of -3.65$\pm$1.55 eV/atom, -4.00$\pm$1.35 eV/atom, and -3.42$\pm$1.33 eV/atom, respectively.

In Fig. \ref{fig:dataset_comparison_including_OMat}b, we plot the force distribution for these three datasets. \datasetname\, demonstrates a slightly higher mean force magnitude of 1.03 eV/\angstrom than MatPES (0.94 eV/\angstrom). OMat contains a notably higher mean force of 1.98 eV/\angstrom, attributable to the more aggressive `rattling', Boltzman sampling, and high-temperature AIMD used by OMat. 

In Fig. \ref{fig:dataset_comparison_including_OMat}c, we plot the pressure distribution for these three datasets. \datasetname\, contains the widest distribution of pressures, with a mean absolute pressure of 13.23 GPa, relative to MatPES and OMat with 4.32 and 4.58 GPA, respectively. This is due to the structures in MatPES and OMat originating from relaxed structures from the Materials project \cite{jain_commentary_2013} and Alexandria \cite{schmidt_crystal_2021, schmidt_machine-learning-assisted_2023}, respectively. \datasetname\,, being initialized from unrelaxed hypothetical structures, hence covers a much broader diversity of pressures.

\clearpage
\subsection*{Molecular Dynamics Stability up to 500 GPa and 3000 K}
\label{sec:md_stability_continued}

\begin{figure}[htb]
	\makebox[\textwidth][c]{\includegraphics[width=0.6\linewidth,trim={0mm 0mm 0mm 0mm},clip]{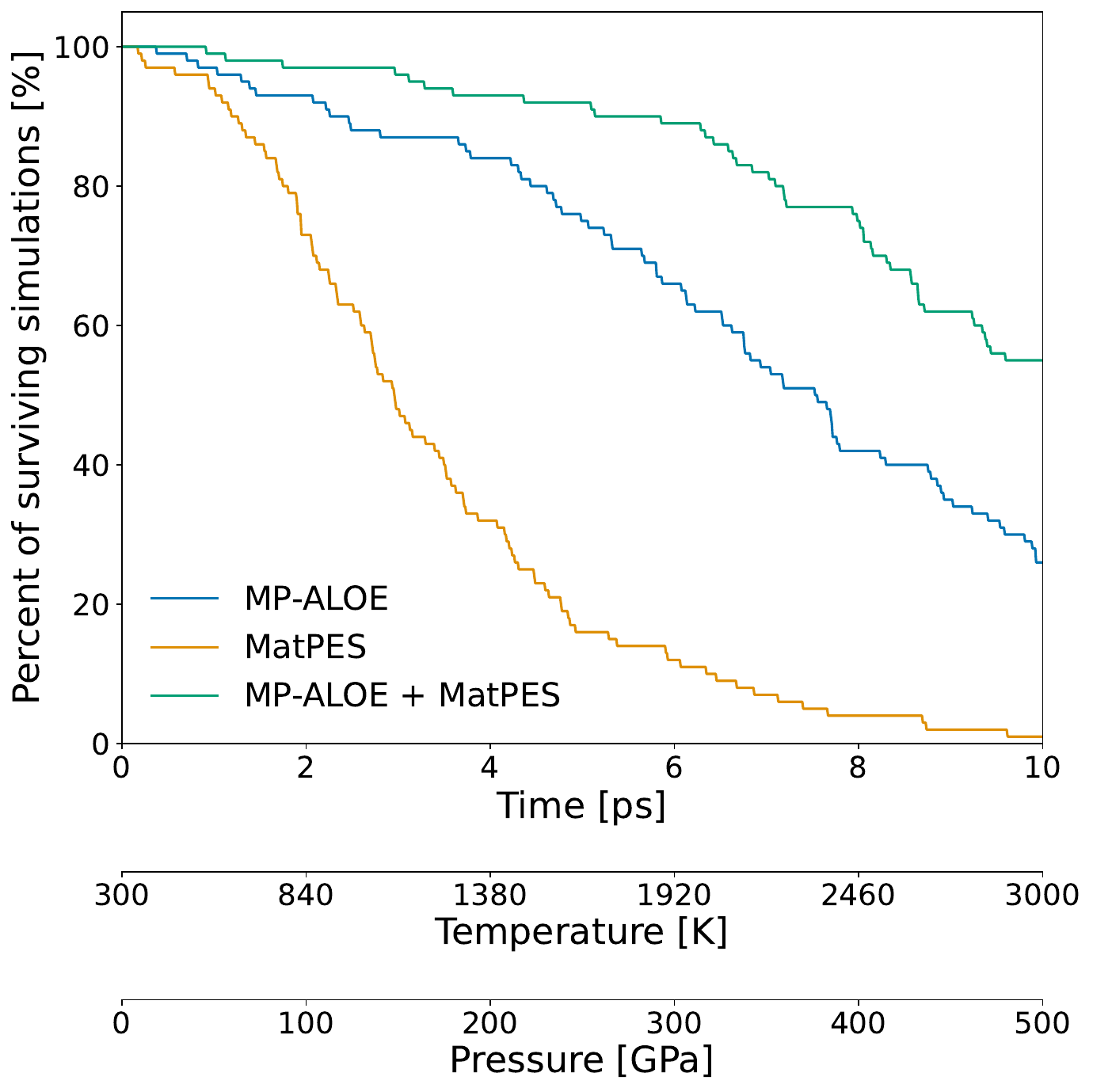}}
	\caption{\textbf{Stability during molecular dynamics simulations}. 100 NPT simulations with a linear temperature ramp from 300 to 3000 K \textit{and} a linear pressure ramp from 0 to \textbf{500} GPa. Tasks and the random mixture structures were sourced from \texttt{MLIP Arena} \cite{chiang_mlip_2025}. The y-axis plots the percentage of surviving runs with a valid MD trajectory at a given timestep. Here, a given trajectory frame is deemed ``valid" simply if the energy is within 100 eV/atom of the initial frame, the maximum kinetic energy for an atom is less than 100 eV, and the cell volume has changed by less than a factor of 10.}
	\label{fig:MD stability_500GPa} 
\end{figure}

Here we plot the original NPT sub-task from \texttt{MLIP Arena}, wherein 100 random mixture structures are simulated for 10 ps with a linear 300 to 3000 K temperature scaling \textit{and} linear pressure scaling from 0 to \textbf{500} GPa. Results are shown in Fig. \ref{fig:MD stability_500GPa}. Here, \datasetname\, significantly outperforms MatPES, which completed 69.9\% and 34.1\% of scheduled timesteps, respectively. This can likely be attributed to the greater sampling of larger pressures in \datasetname, as shown in Fig. 2c from the main text. The combined \datasetname\, + MatPES further improves on this, completing 85.9\% of scheduled timesteps.

Further details on this benchmark task and the random mixture structures used can be found in Ref. \cite{chiang_mlip_2025}.

\clearpage

\subsection*{Validation of hypothetical structures}

As the hypothetical structures generated in this work are not guaranteed to be conventionally charge balanced, nor to contain bonding arrangements known to be well-described by r$^2$SCAN, we have performed a set of minimal validation checks to ensure the structures are not too far from the realm of the possible.

A primary concern is the realistic nature of the oxidation states assigned to the structures.
Automatic assignment of oxidation states is nontrivial, as various bonding arrangements can change the oxidation states a human would assign.
For this test, we collected all possible oxidation states that were (presumably) manually assigned to elements in structures in the Inorganic Crystal Structure Database (ICSD) \cite{zagorac_recent_2019}, and removed ones with less than 1\% representation.
To reflect the fact that metals and monatomic solids should be permitted to take on zero-valued oxidation states, we excluded zero-valued oxidation states (which have representation for all elements) for any multi-element solid containing one of the following elements: H, B, C, N, O, F, Si, P, S, Cl, Ge, As, Se, Br, Sb, Te, I.
We then assigned oxidation states in order of decreasing commonality until a charge balanced configuration could be assigned.
If no charge balanced configuration could be assigned, a structure failed this test.

Second, it is unclear when the overlap of pseudo-potential core radii becomes significant.
This is a quite common situation, especially in higher density structures, which in the case of softer pseudopotentials, can lead to 10 meV/atom errors in energies \cite{furthmuller_softening_2000}.
To assess whether solids had a significant overlap of pseudopotential core radii, we identified each site's nearest neighbor and checked if their separation exceeded the outermost pseudopotential cutoff radius (``RCORE'' in VASP's POTCAR).
We log the number of structures with at least one set of overlapped pseudo-core radii.

Results are shown in Table \ref{tab:sanity_checks}, wherein these three datasets pass both checks at comparable rates. Overall, while MP-ALOE contains a higher proportion of likely charge-imbalanced structures than the Materials Project or MatPES, it contains fewer geometries which sample interatomic distances that are too short for the pseudopotential approximation to hold. 

\begin{table}[h]
    \caption{
        \textbf{Percentages of structures which pass a set of physically motivated checks for a few DFT datasets.}
        We consider the entire MP database, the r$^2$SCAN subset of MatPES, and the current \datasetname\, dataset.
        The validity of the pseudopotential approximation breaks down when pseudo-core regions overlap to a significant degree \cite{furthmuller_softening_2000}, and moreover terms in the pseudopotential total energy involving overlap of pseudo-core densities are neglected \cite{kresse_ultrasoft_1999}.
        Thus structures where at least one nearest neighbor separation is less than the sum of pseudopotential core radii may be errant.
        In the case of VASP's pseudopotentials, we use the outermost pseudopotential core radii, ``RCORE.''
        However, it can be seen that all three databases contain many structures with overlapped pseudo-core regions.
        Further insight is needed to glean what degree of overlap leads to marked errors.
        We also attempted to charge-balance structures in the three databases, using oxidation states with at least 1\% representation in ICSD. Higher percentages are better.
    }
    \begin{tabular*}{\linewidth}{@{\extracolsep{\fill}}lrrrr} 
        \hline
        Dataset & No pseudo-core overlap & Realistic oxidation states & Both metrics & Dataset size \\ \hline
        MP & 41.0\% & 73.7\% & 30.4\% & 394954 \\
        MatPES & 35.2\% & 74.9\% & 27.7\% & 387897 \\
        MP-ALOE & 57.8\% & 55.8\% & 29.9\% & 909792 \\
        \hline
    \end{tabular*}
    \label{tab:sanity_checks}
\end{table}

\end{document}